\documentclass[doublecol]{epl2} 
\usepackage{amsmath}
\usepackage{graphicx}
\usepackage{graphics}
\usepackage{graphicx}
\usepackage{epsfig}
\usepackage{color}
\usepackage{amsmath}
\usepackage{amssymb}
\usepackage{amsfonts}
\usepackage{amsthm}

\title{A Low Order Theory of Arctic Sea Ice Stability}
\shorttitle{Theory of Sea Ice Stability} 

\author{Woosok Moon\inst{1} \and J.~S. Wettlaufer\inst{1,2,3} }
\shortauthor{W. Moon and J.S. Wettlaufer}

\institute{                    
  \inst{1} Department of Geology \& Geophysics - Yale University, New Haven, CT, 06520-8109, USA\\
  \inst{2} Department of Physics \& Program in Applied Mathematics - Yale University, New Haven, CT, 06520-8109, USA\\
  \inst{3} NORDITA,  Roslagstullsbacken 23, SE-10691 Stockholm, Sweden
}

\abstract{
We analyze the stability of a low-order coupled sea ice and climate model and extract the essential physics governing the time scales of response as a function of greenhouse gas forcing.  Under present climate conditions the stability is controlled by longwave radiation driven heat conduction.  However, as greenhouse gas forcing increases and the ice cover decays, the destabilizing influence of ice-albedo feedback acts on equal footing with longwave stabilization.  Both are seasonally out of phase and as the system warms towards a seasonal ice state these effects, which underlie the bifurcations between climate states, combine exhibiting a ``slowing down'' to extend the intrinsic relaxation time scale from $\sim $ 2 yr to 5 yr. }

\begin{document}

\maketitle

\section{Introduction}

Earth's climate is controlled by solar energy input, longwave radiative output and the internal redistribution of energy by the atmosphere and the ocean.  Of the 5 PW excess energy input at low latitudes about two thirds is transferred poleward by the atmosphere and the remainder by the ocean.  One can ascribe an energy budget to each subsystem and among these the polar regions are of particular interest due to their strong influence on the planetary albedo.  The ice-covered polar oceans are understood to be a sensitive diagnostic subsystem for changes in climate, because the contact between the atmosphere and the ocean is mediated by a thin veneer of sea ice that has an albedo about a factor of three larger than that of the ocean.  The ice-albedo feedback is always positive amplifying perturbations in ice coverage and tending to drive large scale glaciation or deglaciation \cite{Saltzman:2002}.  
Indeed, the magnitude of the seasonal areal change of sea ice is 10-20 M km$^2$  in the Arctic and Antarctic respectively, and hence constitutes one of the largest variations in surface albedo on seasonal time scales as does any component of the climate system.  While basic physics and paleoclimate data show that large increases in greenhouse forcing lead to the decay or vanishing of the ice cover, the state of the art numerical global climate models have not made reliable projections of the substantial rate of decay of the minimum annual ice coverage observed by satellites \cite{OneWatt}.    

The following facts motivate this study.  ($i$) There are only about thirty years of satellite observations, during the last decade of which the ice cover has decayed rapidly \cite{OneWatt}.  ($ii$) Although we understand the leading order physics governing the system, the state of the art global climate models do not reliably capture these observations.  ($iii$) We cannot use measurements to close the energy budget with the resolution needed to account for the observed ice loss \cite{OneWatt}.  Thus, we are emboldened to appeal to a quantitative stability analysis of a qualitative, but observationally consistent, theoretical treatment to assess the nature and the characteristic times scales of the transitions we may expect.  

Recently Eisenman and Wettlaufer \cite{EW09} (here referred to as EW09) developed a low-order theory that describes the seasonal evolution of the energy state in the Arctic Ocean.  The theory couples the poleward atmospheric heat transport to the surface radiative heat balance through a gray body atmosphere and to the base of the ice via ocean heat flux.  Over climatological time scales the evolution of the ice state is solved as a two-moving boundary problem; perennial ice grows or thins seasonally according to the energy flux balances across its thickness.  As the greenhouse forcing increases ice is lost in summer but returns in winter.  Finally, under further warming this seasonal state gives way to a perennially ice free state via a saddle-node bifurcation.  In analogy with a first order phase transition, it is basic to such a bifurcation that a hysteresis exists between the two perennial states; in this case ice covered and ice free.  Thus, the perennial ice cover only returns after the climate cools to a second bifurcation point where the perennially ice-free Arctic would be sufficiently cold---colder than the original climate state at which the ice disappeared---to freeze.  Here, we analyze the stability of the seasonal steady states of this model to extract the principal stabilizing and destabilizing effects and their associated response time scales.

\section{Response Theory of the Dynamical Model\label{sec:stability}}

Here, we summarize the theory of EW09 and refer the reader to their Appendix for the full derivation.  The state variable $E$ is the energy (with units W m$^{- 2}$ yr) stored in sea ice as latent heat when the ocean is ice-covered or in the ocean mixed layer as sensible heat when the ocean is ice-free, viz., 
\begin{equation}
E\equiv 
\begin{cases}
-L_i h_i & E<0 \textrm{  [sea ice]}\\
c_{ml}H_{ml}T_{ml} & E \ge 0 \textrm{   [ocean]} 
\end{cases},
\label{eq:sum0}
\end{equation}
where $L_i$ the sea ice latent heat of fusion, $h_i$ its thickness, $c_{ml}$ is the specific heat capacity of the ocean mixed layer, $H_{ml}$ is its depth and $T_{ml}$ its temperature. Ignoring colligative effects,
the temperature $T(t,E)$, determined by energy balance across the layer,  is measured relative to the freezing point ${T}_{m}$ as
\begin{align}
 T(t,E) = -{\cal{R}}\left[\frac{F_{D}(t)}{k_i L_i/E - F_T(t)}\right],  
 \label{eq:T}
\end{align}
where the ramp function is ${\cal R}(x \ge 0) = x$  and ${\cal R}(x<0) = 0$, the thermal conductivity of the ice is $k_i$, and the radiative quantities $F_{D}$ and $F_T(t)$ are discussed below.  

The Beer-Lambert law of exponential attenuation of radiative intensity with depth in a medium requires a treatment of the dependence of the surface albedo with $E$.  While this is done with parsimony it is physically realistic; one uses a characteristic ice thickness $h_{\alpha}$ for the extinction of shortwave radiation as follows
\begin{align}
 \alpha(E)=\frac{\alpha_{ml}+\alpha_i}{2}+\frac{\alpha_{ml}-\alpha_i}{2} \text{tanh}\left(\frac{E}{L_i h_{\alpha}}\right),
 \label{eq:alpha}
\end{align}
which controls the fraction, $1-\alpha(E)$, of the incident shortwave radiation $F_S(t)$ absorbed by the ice \cite{EW09}.  

The evolution of the state of the ice (or ocean) cover is determined by the balance of radiative and sensible heat fluxes at the upper surface, $F_{D} - F_T(t) T(t,E)$, the upward heat flux from the ocean $F_B$, and the fraction of ice exported from the domain $v_0 \mathcal{R}(-E)$ ($\sim 10\%$ yr$^{-1}$) through a first order nonautonomous energy balance model as
\begin{align}
 \frac{dE}{dt}=f(t,E),
\label{eq:DS}
\end{align}
with
\begin{align}
 f(t,E)=F_{D}-F_T(t)T(t,E)+F_B+v_0 \mathcal{R}(-E), 
\end{align}
where 
\begin{equation}
F_{D}(t,E) \equiv  \left[1-\alpha(E)\right] F_S(t) -F_0(t) + \Delta F_0. 
\label{eq:FD}
\end{equation}
The term $F_{D} - F_T(t) T(t,E)$ is thought of as the difference between the incoming shortwave radiation at the surface $\left[1-\alpha(E)\right] F_S(t)$ and the outgoing longwave radiation
($\propto T^4$), augmented here by sensible and latent heat fluxes and an additional amount associated with greenhouse gas forcing $\Delta F_0$.
The Stefan-Boltzmann equation for outgoing longwave radiation is linearized in the deviation of the surface temperature from the freezing point as $F_0(t) + F_T(t) T(t,E)$ which is known to be a reasonable approximation \cite{AST92, EW09}.  Finally, the seasonally varying values of $F_0(t)$ and $F_T(t)$ are determined from an atmospheric model incorporating observations of Arctic cloudiness, atmospheric transport from lower latitudes and the meridional temperature gradient \cite{EW09}. 

The seasonal steady states ($E_\text{S}$), or periodic points, of Eq. (\ref{eq:DS}) were described heuristically in the introduction and examples are shown Fig. 3 of EW09, wherein sensitivities to greenhouse forcing and other parameters were studied.  Here we assess the nature of the stability of these periodic points and their relaxation time scales. 

Consider $E(t)=E_\text{S}(t)+\xi(t)$, where $|\xi(t)|  \ll  |E_\text{S}(t)|$ and thus
\begin{eqnarray}
 \frac{dE}{dt} & = &\frac{dE_\text{S}}{dt}+\frac{d\xi}{dt} \nonumber \\ & = & f(t,E_\text{S}+\xi)\simeq f(t,E_\text{S})+{\left.\frac{\partial f}{\partial E}\right|_{E=E_\text{S}}}\xi.
\end{eqnarray}
Now, because ${dE_\text{S}}/dt=f(t,E_\text{S})$ we have 
\begin{align}
 \frac{d\xi}{dt} = {\left.\frac{\partial f}{\partial E}\right|_{E=E_\text{S}}}\xi \equiv a(t)\xi,
\end{align}
which has an exact solution
\begin{align}
 \xi(T) = \xi(0)\text{exp}\left[\int_{0}^{T} \! a(s) \, ds\right] \equiv ~\xi(0) \text{e}^\gamma ,
\end{align}
written here for the annual cycle of the system $T$.  Therefore, an unstable (stable) periodic point of Eq. (\ref{eq:DS}) has $\gamma > 0$ ($\gamma < 0$) with
a perturbation relaxation time scale of divergence from (convergence to) a particular periodic point determined by the competing effects embodied in $a(t)$, itself depending on whether there is ice, $E<0$, $F_{D}(t)<0$, or ocean, $E \geq 0$.  If $F_{D}(t)\geq 0 $ the ice ablates downward from the surface while $T(t,E)$ is pinned at zero.  Accordingly, we examine the structure of $a(t)$ in the two thermodynamic regimes; the ice covered and ice free states. 

\subsection{ Ice Covered State: $E<0$, $F_{D}(t)<0$}

Generally we find 
\begin{align}
 a(t)= - \left.\frac{\partial \alpha}{\partial E}\right|_{E=E_\text{S}} F_S(t) - \left.\frac{\partial T(t,E)}{\partial E}\right|_{E=E_\text{S}}F_T(t)-v_0, 
\label{eq:Gena(t)}
\end{align}
in which 
\begin{align}
 \left.\frac{\partial \alpha}{\partial E}\right|_{E=E_\text{S}} = \frac{\alpha_{i}-\alpha_{ml}}{2L_i h_{\alpha}} \left[1-{\tanh}^2 \left(\frac{E_\text{S}}{L_i h_{\alpha}}\right) \right] \equiv \frac{a_{IA}(t)}{F_S(t)},
 \label{eq:aIA(t)}
\end{align}
and 
\begin{align}
\left.\frac{\partial T(t,E)}{\partial E}\right|_{E=E_\text{S}}= &-\frac{a_{IA}(t)}{-k_i L_i/E_\text{S}+F_T(t)} \nonumber \\
 &+\frac{F_{D}(t, E_\text{S})}{[-k_i L_i/E_\text{S} + F_T(t)]^2}\frac{k_i L_i}{{E_\text{S}}^2}\nonumber \\
 & \equiv \frac{a_{AR}(t)+a_{LW}(t)}{F_T(t)}.
\label{eq:Tderiv}
\end{align}
Therefore, we write Eq. (\ref{eq:Gena(t)}) as 
\begin{align}
 a(t)=a_{IA}(t)+a_{AR}(t)+a_{LW}(t)+a_{EX}(t),
\label{eq:a(t)}
\end{align}
where  $a_{IA}(t)$ describes the ice-albedo feedback, $a_{AR}(t)$ and $a_{LW}(t)$ the albedo and longwave responses, and $a_{EX}(t) \equiv - v_0$ the ice export, each of which we describe below.
Note that the derivative of $T(t,E)$ with respect to $E$ in Eq. (\ref{eq:Tderiv}) depends on the sign of $F_{D}(t)$. When $F_{D}(t) \ge 0$ and the ice is ablating ${\partial T(t,E)}/{\partial E}\mid_{E=E_\text{S}} = 0$ so that $a(t)=a_{IA}(t)+a_{EX}(t)$.  

\subsection{ Ice-Free State: $E \ge 0$, $F_{D}(t) \ge 0$}

The response of the system is limited by the radiative balance over the ocean mixed layer and hence we have
\begin{align}
 a(t) = -\frac{F_T(t)}{c_{ml}H_{ml}}.
\end{align}

\section{Dissecting the Stability Parameters; $a(t)$ \& $\gamma$ \label{sec:gamma}}

First we describe the climatological evolution of the direct response rate $a(t)$, for it underlies the seasonal evolution of the ice state.  Next in more detail we analyze  the integrated influence of those dynamics on the perturbations to the system viz., $\gamma$.  Note that the dominant underlying physics influencing stability is laid bare by this simple analysis, but is included in  more complex numerical models (see e.g., \cite{OneWatt} and refs therein).

\begin{figure}[!hbtp]
\centering
\includegraphics[angle=0,scale=0.46,trim= 0mm 0mm 0mm 0mm, clip]{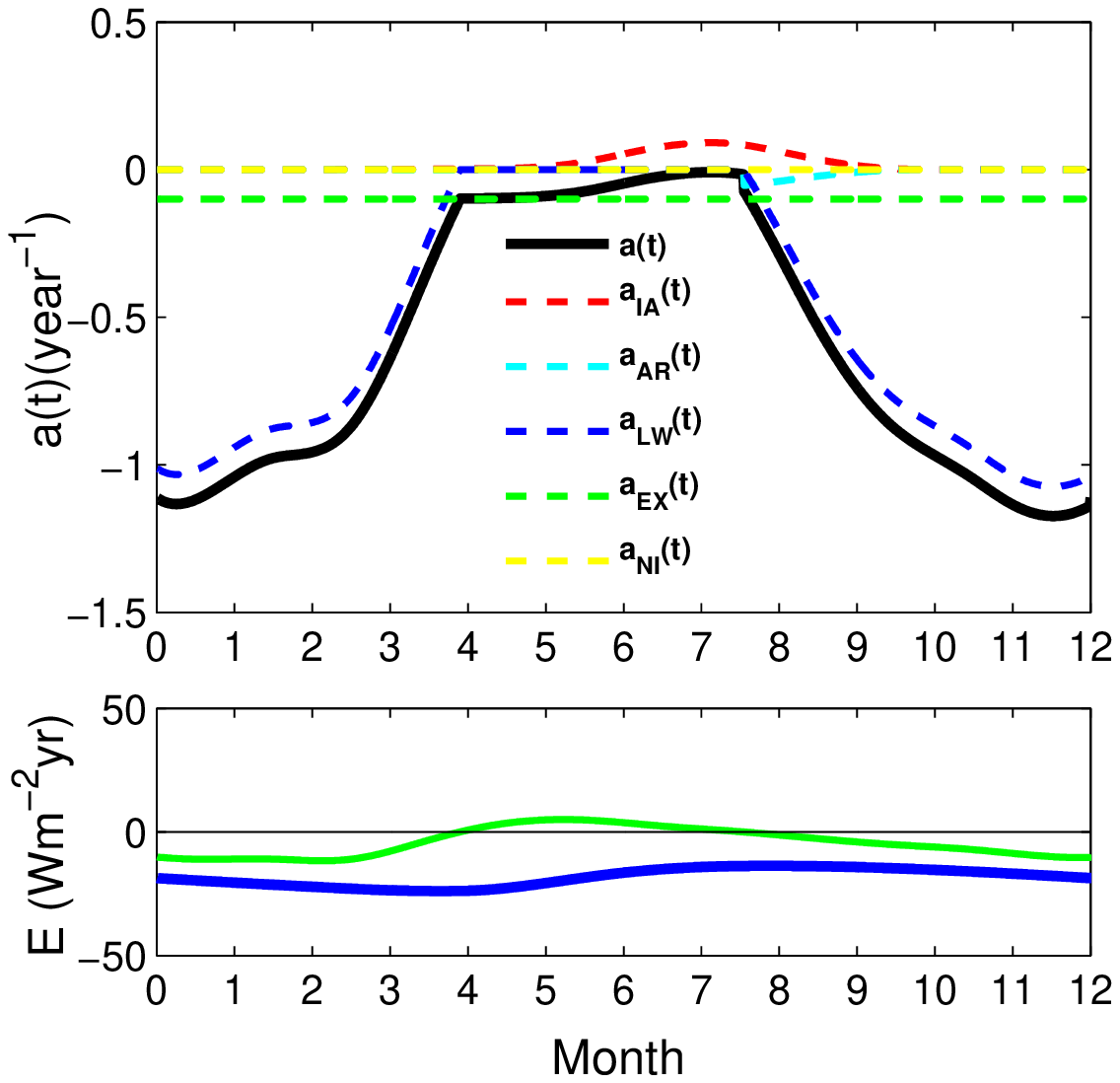}
\includegraphics[angle=0,scale=0.46,trim= 0mm 0mm 0mm 0mm, clip]{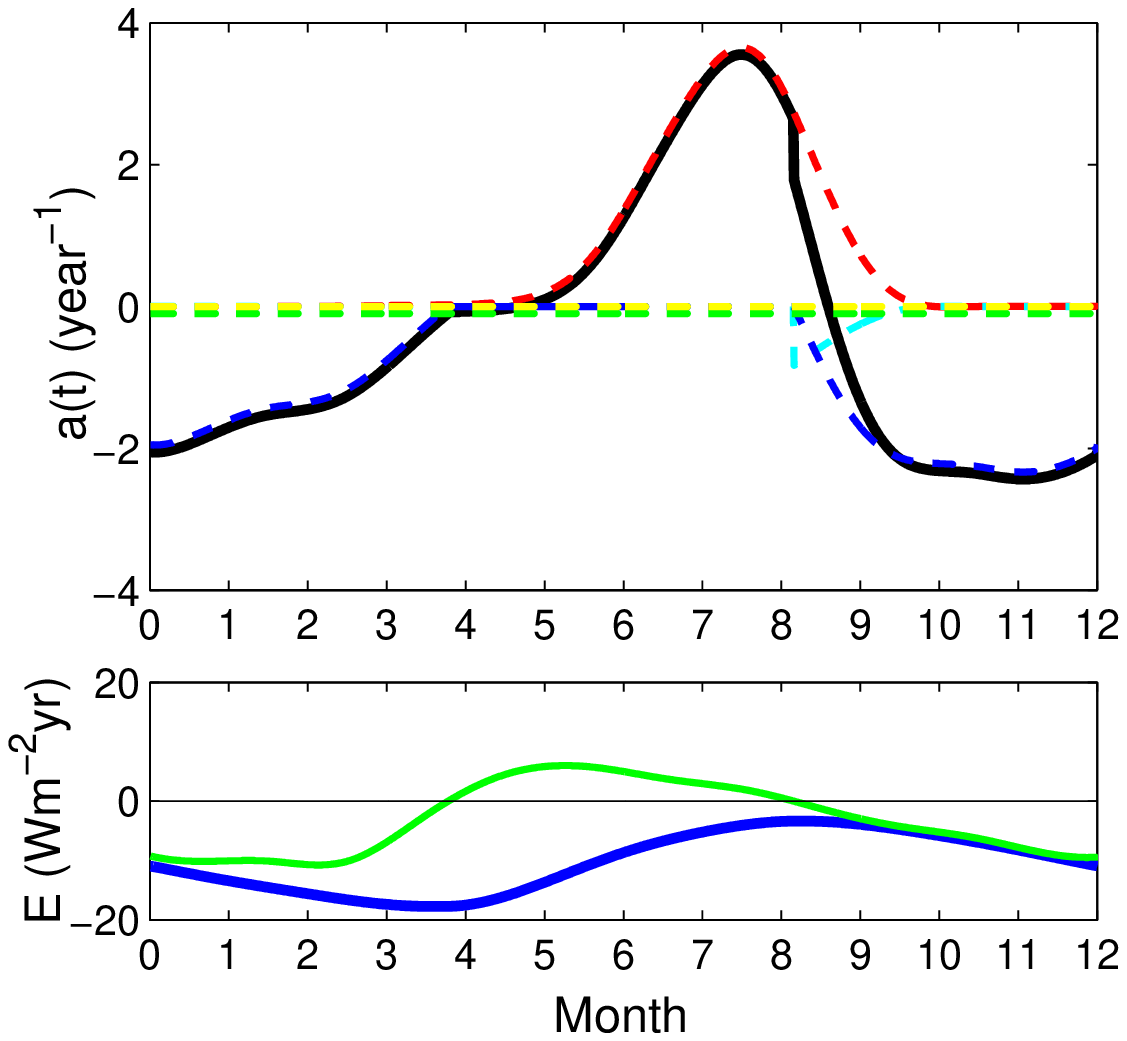}
\includegraphics[angle=0,scale=0.46,trim= 0mm 0mm 0mm 0mm, clip]{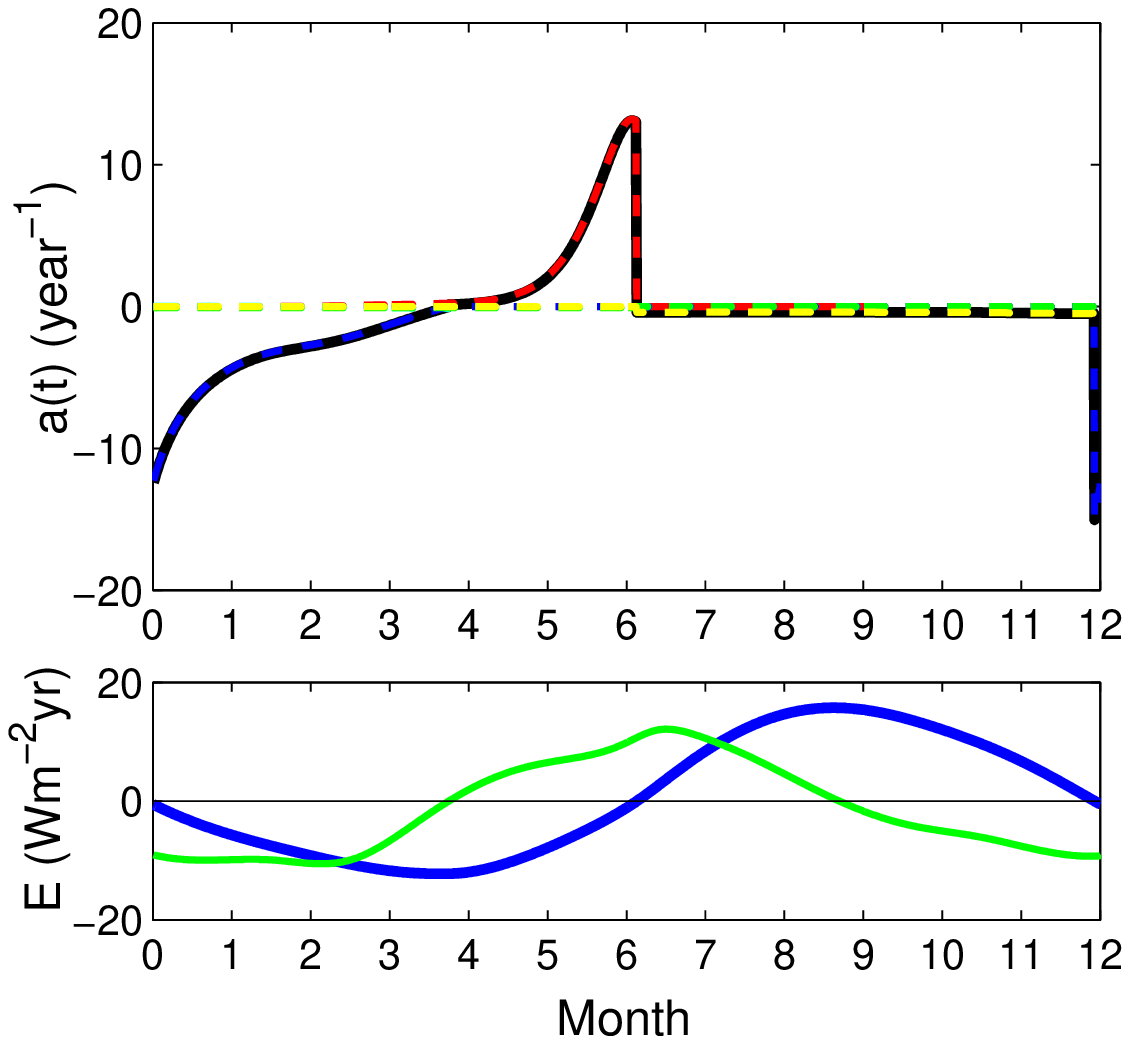}
\caption{The response rate $a(t)$ of Eq. (\ref{eq:a(t)}) and all of the contributions defined therein as a function of month for three values of of greenhouse gas forcing $\Delta F_0$ of 10 (top), 19 (middle) and 21 (bottom)  W m$^{-2}$, along with the evolution of the energy $E_\text{S}(t)$ (lower blue curves) described by Eq. (\ref{eq:DS}) and the radiative forcing $F_{D}(t)$ (lower green curves) as given by Eq. (\ref{eq:FD}). The solid black curve is the net response rate $a(t)$ and the individual contributions are denoted by the subscripts of Eq. (\ref{eq:a(t)}).  We see the strong seasonal dependence of the dominant contributions to the stability; destabilizing ice-albedo feedback ($a_{IA} > 0$) and the stabilizing influence of heat conduction driven by radiative loss at the ice surface ($a_{LW}< 0$ ).}
\label{fig:a(t)}
\end{figure}

The evolution of the response rate  $a(t)$ and its components as described by Eq. (\ref{eq:a(t)}) are shown in Figure \ref{fig:a(t)} along with the seasonal steady state solution $E_\text{S}(t)$ (blue lower curve) described by Eq. (\ref{eq:DS}) and $F_{D}(t)$ (green lower curve) given by Eq. (\ref{eq:FD}).
The net radiative forcing and the surface temperature determine which components of $a(t)$ are dominant. During winter, when $F_{D}(t)$ is negative, $a_{LW}(t)$ dominates, whereas during summer $a_{IA}$ dominates.  As greenhouse forcing $\Delta F_0$ increases the dynamics of $a(t)$ change and we see that when $\Delta F_0$  = 19.0 W m$^{-2}$, although the steady 
state solution still represents a perennial ice state, during summer sea ice becomes sufficiently thin  that sea ice albedo feedback 
nearly drives an instability in the ice cover.   Indeed, the principal difference relative to the 
$\Delta F_0$ = 10.0 W m$^{-2}$ case is the magnitude of $a_{IA}(t)$ during summer which controls $a(t)$; starting in April the ice starts to thin and by May its  thickness lies in the range where the ice albedo feedback controls $E_\text{S}(t)$. However, by September $F_{D}(t)$ turns negative and thin ice grows rapidly.  By $\Delta F_0$ = 21 W m$^{-2}$ the seasonal ice state appears.   Whilst these dynamics provide a detailed picture of the seasonal variation of the stability of the system and their influence on the seasonal steady state solutions $E_\text{S}(t)$ as greenhouse forcing increases, it is the seasonally averaged contributions embodied in $\gamma$ that most succinctly describe the overall stability.  

\begin{figure}[!hbtp]
\centering
\includegraphics[angle=0,scale=0.67,trim= 0mm 0mm 0mm 0mm, clip]{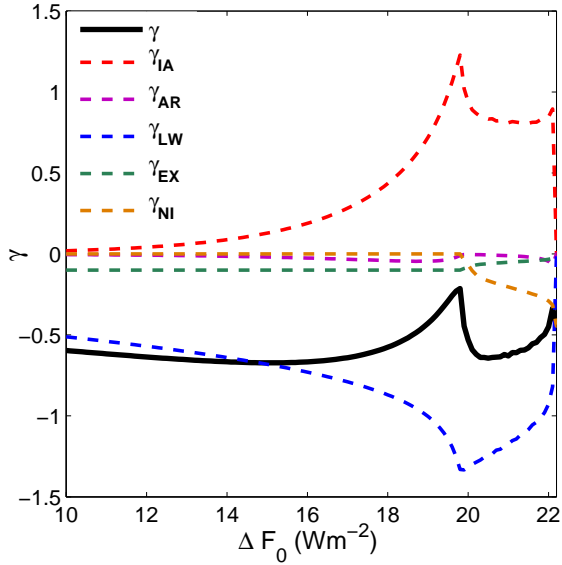}
\caption{The response of the ice cover $\gamma$ to perturbations as a function of greenhouse gas forcing $\Delta F_0$. When $\gamma < 0$ ($\gamma > 0$) the periodic points $E_\text{S}$ of Eq. (\ref{eq:DS}) are stable (unstable), and their relaxation time scale is $|1/\gamma|$.
The solid black curve is the entire response which for small $\Delta F_0$ is dominated by the longwave contribution  ($\gamma_{LW} < 0$; blue dashed curve).  As the ice thins the ice albedo feedback ($\gamma_{IA} > 0$) figures more prominently (red dashed curve) and the stability is dominated by the competition between the destabilizing ice-albedo feedback and the stabilizing influence of heat conduction driven by radiative loss at the ice surface. The local maximum at $\Delta F_0 = 19.8$ W m$^{-2}$ is the transition from the perennial to the seasonal (winter only) ice state and the second increase at $\Delta F_0 = 22.2$ W m$^{-2}$ denotes the saddle-node bifurcation to a seasonally ice free state. The small differences between these thresholds and EW09 are 
because we use $\alpha(E)$ rather than  $\alpha_i$ in  Eq. (\ref{eq:T}). }
\label{fig:gamma}
\end{figure}

Following the same notation as in Eq. (\ref{eq:a(t)}) Figure \ref{fig:gamma} shows $\gamma$ and its constituents as a function of the greenhouse forcing $\Delta F_0$.   The range of $\Delta F_0$ shown spans the perennial and seasonal ice states of the system.  In the following paragraph we summarize the principal results, and then discuss the contributions to $\gamma$ in turn.   

For greenhouse forcing up to $\Delta F_0 \sim$ 16 W m$^{-2}$, the perennial ice is sufficiently thick that the ice-albedo feedback plays a minor role and longwave radiative control of ice growth in winter dominates the stability with a rather rapid relaxation time of a few years.  As the climate warms and the ice thins the stabilizing influence of heat conduction acts on a much more rapid relaxation time scale (thin ice grows more rapidly than thick ice \cite{Stefan:1891}) but so too does the destabilizing ice-albedo feedback governed by Eq. (\ref{eq:aIA(t)}).  These competing effects lead to an overall {\em increase} in the relaxation time with greenhouse forcing of $\sim$ 5 yr until $\Delta F_0 \sim$ 19.8 W m$^{-2}$, ice is lost in summer, and reentrant longwave control of ice growth in winter dominates the rapid stability of this state of the system.  Finally, when $\Delta F_0 \gtrsim$ 22.2 W m$^{-2}$, the ice is too thin to thwart the ice-albedo feedback, it does not recover in the polar night, and is thus lost entirely.  

The first contribution, $a_{IA}(t)$, describes the fluctuation in the surface shortwave radiative flux associated with a perturbation in ice thickness.  
The fundamental operation of the ice-albedo feedback is laid bare by the fact that $a_{IA}(t)$ is always positive; amplifying {\em all perturbations} in ice thickness.  Eq. \ref{eq:aIA(t)} is rather transparent in the display of the albedo contrast as a function of the ice thickness. Although $F_S(t) = 0$ during winter, the annual manifestation of the ice-albedo feedback is $\gamma_{IA}=\int_0^T \! a_{IA}(s) ds > 0$ and, as  $\Delta F_0$ increases, this drives the transition to both the seasonal ice and ice-free states.  Such behavior is due to the fact that $\gamma_{IA}$ only begins to play a controlling role in the state of the system when $h \lesssim h_{\alpha}$.  Importantly, a value for $h_{\alpha}$ of 0.5 m is realistic given the optical properties of sea ice.  However, an unphysical value of even half of this results in unrealistic transitions in the state of the system; albedo is a material property.  

The Stefan-Boltzmann law underlies the principal stabilizing influence, understood from rudimentary--radiative steady state--studies of the greenhouse effect.  Increasing the flux $F_{D}(t)$ increases the surface temperature $T(t,E)$ and hence the outgoing longwave flux via the Stefan-Boltzmann law.  In this coupled model it is prudent to take care in interpreting this feedback.  Here, any change in $T(t,E)$ has an immediate influence on the heat conduction through the ice but the atmospheric meridional heat flux is proportional to the difference between the ice surface temperature and that at the lower latitudes.  For example, when $F_S(t)$ increases during early summer, both $T(t,E)$ and the associated outgoing longwave radiance increase, but the atmospheric meridional heat flux decreases.  Thus, to leading order these two effects have the same sign.  Whereas, in winter $F_S(t) = 0$ and any anomalously thin ice grows more rapidly due to the basic tenets of heat conduction§
 driven by longwave loss at the surface.  
Combined these processes are captured by $a_{AR}(t)$ and $a_{LW}(t)$.  It is evident that because $a_{AR}(t) = a_{IA}(t)$/$[-k_i L_i/E_\text{S}+F_T(t)]$, this term describes the stabilizing effect of heat conduction and outgoing longwave radiance weighed against the perturbation in incoming shortwave radiance associated with the fluctuation in the surface albedo.  Whence, we call this ``albedo response''. Whilst $\gamma_{AR} < 0$, its magnitude is small because $T(t,E) = 0$ and hence $a_{AR}(t) = 0 $ when the ice is ablating during summer, and $a_{AR}(t) = 0 $ during winter because $F_S(t) = 0$.   Therefore, the stabilizing action of $a_{AR}(t)$ is confined to the spring and fall. 

The principal influence embodied in $a_{LW}$ is the growth of ice by longwave cooling during winter; perturbations associated with ice-albedo feedback when $F_S(t) >0$ are strongly compensated for during the polar night. The thinner the ice at the end of the summer the greater, and more rapid, the stabilizing growth response. While the ice  ablates in summer $T(t,E) = 0$ and hence $a_{LW}(t)$ vanishes.  However, it describes the dominant stabilizing effect in the system insuring that $\gamma_{LW} < 0$ with $|\gamma_{LW}|$ increasing as  $\Delta F_0$ increases and the ice thins.   

Finally, the rate of sea ice export out of the Arctic Ocean, controlled by wind and ocean forcing, is represented by $\gamma_{EX}$, and when the ice vanishes the  
longwave response is shown as $\gamma_{NI}$. 

The dominant competition determining the stability of the periodic points of the theory, and the associated relaxation rate as $\Delta F_0$ increases, is between the ice-albedo feedback $a_{IA}(t)$ and the nonlinear response of ice growth associated with longwave loss in winter $a_{LW}(t)$.  These two effects act in opposite seasonality and hence provide an intuitive picture for how one state can emerge from another as the persistence of the ice cover is enhanced/suppressed due to fluctuations in forcing.  Moreover, once $\Delta F_0$ has increased beyond about 19.8 W m$^{-2}$ this competition demonstrates the origin of the robustness of winter ice.  Minor manipulation affords some insight into the main stabilizing effect viz., 
\begin{align}
{a_{LW}(t)}^{-1}
=&\frac{F_T(t)\tilde{T}}{k_i \tilde{T}/h_\text{S}}\frac{E_\text{S}}{F_{D}}\left[1 + \frac{k_i}{{h_\text{S}} F_T(t)}\right]^{2}
\label{eq:aLW}
\end{align}
where $\tilde{T}$ is the perturbation in surface temperature and $h_\text{S}$ is the steady state ice thickness.  
This form of $a_{LW}(t)$ displays the stabilizing effects $\frac{F_T(t)\tilde{T}}{k_i \tilde{T}/h_\text{S}}$ and $\frac{E_\text{S}}{F_{D}}$;  the longwave heat loss controlling heat conduction, and the fraction of radiative forcing stored as latent heat in the ice respectively. 
The former effect is basic to heat conduction; for a given longwave radiative forcing thinner ice grows more rapidly than thicker ice.  Thus, as the climate warms and $\Delta F_0$ increases, $h_\text{S}$ decreases leading to a faster response for the stabilization of perturbations.  Considering annual mean values (see Table 1 of EW09) for demonstrative estimates $\frac{F_T(t)\tilde{T}}{k_i \tilde{T}/h_\text{S}}\approx 4.2$  for 3 m thick ice under no greenhouse forcing ($\Delta F_0$=0) and about 1.4 for 1 m thick ice with $\Delta F_0$ = 19.8 W m$^{-2}$ (where perennial ice is lost).  Inspection of Eq. (\ref{eq:FD}) shows that the shortwave absorption at the surface decreases as the ice thickens, rapidly saturating for $h \gtrsim h_{\alpha}$ and, for a given ice thickness, increases with $\Delta F_0$.  For the same conditions given above, $\frac{E_\text{S}}{F_{D}}$ decreases from about 0.5 to 0.3 yr  with increasing $\Delta F_0$.  
Taken together this demonstrates the contributions to the more rapid stabilizing growth response of thin ice in a warming climate as $|{a_{LW}(t)}^{-1}|$ decreases from $\sim$ 3.5 to 1.2 yr.  

\section{Conclusions\label{sec:conclude}}

As seen in figure \ref{fig:gamma} the overall stability and the net response rate of the periodic points of Eq. (\ref{eq:DS}) to perturbations depends on all the contributions to $\gamma$, but it is dominated by the competition between the {\em destabilizing} ice-albedo feedback $\gamma_{IA} > 0$ and the {\em stabilizing} influence of heat conduction driven by wintertime radiative loss at the ice surface $\gamma_{LW} < 0$.  The order of magnitude estimates given above for the latter effect demonstrate the general stabilizing behavior that thin ice responds much more rapidly than thick ice to perturbations in radiative forcing.  Because ice with $h \lesssim h_{\alpha}$ absorbs more shortwave radiation than does thick ice (Eq. \ref{eq:alpha}), the destabilizing ice-albedo feedback also operates much more rapidly as $\Delta F_0$ increases and the ice thins.  Thus as $\Delta F_0$ increases while these principal, seasonally out of phase, contributions to the competition which governs the stability of the system ($\gamma_{IA}$ and $\gamma_{LW}$) each operate on shorter time scales, their difference decreases thereby producing a collective state that responds more slowly ($\sim$ 5 yr as $\Delta F_0 \rightarrow$ 19.8 W m$^{-2}$).   Whereas under conditions of small $\Delta F_0$ more representative of the present climate, 
we find time scales $\gamma^{-1}$ of $\sim$ 2 yr.  Hence, a recent study of the simulated recovery using an atmosphere-ocean general circulation model is relevant in this regard.  Tietsche et al., (2011) numerically prescribed ice-free summer states at various times during the projection of 21$^{\text{st}}$ century climates and found that ice extent typically recovered within several years.  An important implication of their results is that while the ice-albedo feedback may drive the transition to an ice free summer, the stabilization and recovery of the system can be rapid.  Not only are the recovery processes described in this compact theory ostensibly the same as in the complex treatment in Tietsche et al., \cite{Tietsche:2011} but the stability analysis presented here shows that the time scales are commensurate.  Moreover, the underlying processes and their intrinsic time scales are immediately accessible within our framework.  

We close by noting that such a stability analysis lays the foundation for a rigorous study of the role of stochastic forcing in the climate state of the system  \cite{Saltzman:2002}.  This is because while the transitions between the periodic points of Eq. (\ref{eq:DS}) are essential to understanding the basic physics of the system \cite{EW09}, the nature of these transitions will depend intimately on the response time scales ($\gamma^{-1}$) to fluctuations.  Because of the secular trends in $\gamma_{IA}$ and $\gamma_{LW}$ as the climate warms, the detailed interplay between these competing effects in the presence of noise comprises a separate detailed study.

\acknowledgments
Both authors thank Yale University for support of this research.  WM thanks NASA for a graduate fellowship and JSW thanks the Wenner-Gren Foundation and the John Simon Guggenheim Foundation.


\begin{thebibliography}{0}

\bibitem{Saltzman:2002}
  \Name{Saltzman B.}
  \Book{Dynamical Paleoclimatology: Generalized Theory of Global Climate Change}
  \Publ{Academic Press, San Diego}
  \Year{2002}
  \Page{357}.

\bibitem{OneWatt}
  \Name{Kwok R. \and Untersteiner N.}
  \REVIEW{Phys. Today}{64}{2011}{36}.
  
 \bibitem{EW09}
  \Name{Eisenman I. \and Wettlaufer J.S.}
  \REVIEW{Proc. Natl. Acad. Sci. USA}{106}{2009}{28}.

 \bibitem{AST92}
  \Name{Thorndike A.S.}
  \REVIEW{J. Geophys. Res.}{97}{1992}{9401}.

 \bibitem{Stefan:1891}
  \Name{Stefan J.}
  \REVIEW{Ann. Phys.}{278}{1891}{269}.

 \bibitem{Tietsche:2011}
  \Name{Tietsche S., Notz D., Jungclaus J.H.  \and Marotzke J.}
  \REVIEW{Geophys. Res. Lett.}{38}{2011}{L02707}.


\end{thebibliography}
\end{document}